# General Methodology for developing UML models from UI

Ch Ram Mohan Reddy[1], D Evangelin Geetha[2], KG Srinivasa[2],
T V Suresh Kumar[2], K Rajani Kanth[2]
[1]B M S College of Engineering, Bangalore - 19, India
[2]M S Ramaiah Institute of Technology, Bangalore - 54, India

crams19@yahoo.com, degeetha@msrit.edu, kgsrinivas78@yahoo.com,
tvsureshkumar@msrit.edu, rajanikanth@msrit.edu

## Abstract

*In recent past every discipline and every industry have their own methods of developing products. It may be software development, mechanics, construction, psychology and so on. These demarcations work fine as long as the requirements are within one discipline. However, if the project extends over several disciplines, interfaces have to be created and coordinated between the methods of these disciplines. Performance is an important quality aspect of Web Services because of their distributed nature. Predicting the performance of web services during early stages of software development is significant. In Industry, Prototype of these applications is developed during analysis phase of Software Development Life Cycle (SDLC). However, Performance models are generated from UML models. Methodologies for predicting the performance from UML models is available. Hence, In this paper, a methodology for developing Use Case model and Activity model from User Interface is presented. The methodology is illustrated with a case study on Amazon.com.*

## 1 Introduction

System engineering concentrates on the definition and documentation of system requirements in the early development phase. The preparation of a system design, and the verification of the system as to compliance with the requirements, taking the overall problem into account: operation, time, test, creation, cost and planning, training and support, and disposal [21]. Systems engineering integrates all disciplines and describes a structured development process, from the concept to the production to the operation phase and finally to putting the system out of operation. It looks at both technical and economic aspects to develop a system that meets the users' needs. Hence a new combination of technologies is required to cop up with the new demand.

### 1.1 Graphical User Interface.

Over some period graphical user interface have become increasingly dominant, and design of the external or visible system has assumed increasing importance. This has resulted in more attention being devoted to usability aspects of interactive system and a need for development of tools to support the design of the external system. Model and notations are required for describing user tasks and for mapping these tasks on to user interface.

The primary purpose of task models is to define the activities of the user in relation to the system, as a means of uncovering functional requirements to be supported by the system. Task model focuses on tasks decomposing and/or task flow. A variety of task description methods have been developed, often involving graphical techniques.

Activity diagram, as supported in UML, falls into the latter category. The user of task models in interface design is limited by a lack of tools and techniques to support their definition and weak linkage to visual interface design.

## 1.1 Importance of UML 2.0

In recent past, many researches and software industry use Unified Modeling Language (UML) for the conceptual and logical modeling of any system because of the advantages it has. UML supports both static and dynamic modeling. UML 2.0 is the newer version, which has more features that can be useful for modeling complex systems also [4],[17].

The standard mechanism that UML provides are adaptable itself to a specific method or model, such as constraints and tagged values. we use UML to design RMA processes because it consider an information systems structural and dynamic properties at the conceptual level more naturally than do classic approaches such as Entity- Relationship model. This approach for modeling RMA processes yields simple yet powerful extended UML use case and sequence diagrams that represent RMA properties at the conceptual level.

There are three classifications of UML diagram.

Behavior diagram. A type of diagram that depicts behavioral features of a system or business process. This includes activity, state machine, and use case diagrams as well as the four interaction diagrams.

Interaction diagrams. A subject of behavior diagrams which emphasize object interactions. This includes communication, interaction overview, sequence, and timing diagrams.

Structure diagrams. A type of diagram that depicts the elements of specification that is irrespective of time. This includes class, composite structure, and component, deployment object and package diagrams.

## 1.1 Web Services

The emergence of Web services introduces a new paradigm for enabling the exchange of information across the Internet based on open Internet standards and technologies. Using industry standards, Web services encapsulate applications and publish them as services. These services deliver XML-based data on the wire and expose it for use on the Internet, which can be dynamically located, subscribed, and accessed using a wide range of computing platforms, handheld devices, appliances, and so on.

Since early 2006, Amazon Web Services (AWS) has provided companies of all sizes with an infrastructure web services platform in the cloud. With AWS compute power, storage, and other services such as gaining access to a suite of elastic IT infrastructure services can be requested. Using Amazon Web Services, an e-commerce web site can weather unforeseen demand with ease; a pharmaceutical company can “rent” computing power to execute large-scale simulations; a media company can serve unlimited videos, music, and more; and an enterprise can deploy bandwidth-consuming services and training to its mobile workforce [23]. Amazon Web Services delivers a number of benefits for IT organizations and developers alike, including:

- Cost-effective. The uses of AWS have to pay only for what they use, as they use it, with no up-front commitments. As the Amazon Web Services cloud grows, the operations, management and hardware costs shrink.
- Dependable. AWS is a battle-tested; web-scale infrastructure that handles whatever that has been throwing at it. The Amazon Web Services cloud is distributed, secure and resilient, giving reliability and massive scale.
- Flexible. AWS provides flexibility to build any application, using any platform or any programming model. The resources can be controlled and fit into your application.
- Comprehensive. Amazon Web Services gives you a number of services that can incorporate into your applications. From databases to payments, these services help to build any great applications cost effectively and with less up-front investment.

## 1.1 Transformation of UML models into Performance models

Critical aspects of the quality of a software system are its performance. There are many software engineering methodologies focus on the functionality of the system, while applying a "fix-it-later" principle to software performance aspects. The system is designed to meet its functional requirements, by giving less priority to the performance at the later development stages. This result, lengthy expensive extra hardware, or even redesigns are necessary for the system to meet the performance requirements. Even with the fine-tuning, there is no guarantee that the system performance will be appropriate. Several modeling formalisms have been designed to allow system designers to model the system performance, e.g., queueing networks [6], [14] and Petri Nets [15 ]& [22].

UML model to incorporate the performance related quality of service (QoS) information to allow modeling and evaluating the properties of a system like throughput, utilization, mean response time. So the UML models are annotated according to the Profile for Schedulability, Performance, and Time (SPT) [16] to include quantitative system parameters in the model. Markov models, queuing networks, stochastic process algebras and stochastic petrinet are probably the best studied performance modeling techniques [10]. Markov model is well-developed numerical modeling analysis techniques; it has the ability to preserve the original architecture of the system.

Keeping in view of the above discussion we propose a methodology to transform prototype of User Interface (UI) to UML models. The remaining part of the paper is organized as follows: In Section 2, related work is presented; Section 3 describes the steps involved in the proposed Methodology; The Methodology is illustrated with the Case Study on Amazon.com in Section4. In Section 5, the paper is concluded and the future directions of the work are highlighted.

## 1 Related Work

The evaluation of system specifications early in the software development lifecycle has increasingly gained attention from the software engineering community. Early evaluation of software properties, including non-functional ones, is the important in order to reduce costs in software development before resources have been allocated and decisions have been made. Dependability is one example of an important non-functional property and represents the ability to deliver service that justifiably can be trusted. Discovery of Web Services is of an immense interest and is fundamental area of research in ubiquitous computing. Web Services are Internet-based, distributed modular applications that provide standard interfaces and communication protocols aiming at efficient and effective service integration. A Web Service is defined as a functionality that can be programmatically accessible via the Web [18]. A fundamental objective of Web Services is to enable the interoperability among different software applications that run on a variety of platforms [13], [24]. The interoperation has been enabled by the tremendous standardization effort to describe, advertise, discover and invoke Web Services [3]. Web Services are increasingly being adopted as a framework to access Web-based applications. Most of the proposed composition languages for Web Services are based on XML [1], and although XML-based representations have their advantages as universal representations and exchange formats, they can be difficult to understand and to write for non-XML experts. Thus, the use of a graphical modeling language can be very useful to understand the behavior of Systems.

Performance has been recognized as the most considered aspect for the software system [5].many software performance evolution approaches have been proposed in the literature [19]. UML is the standard for the modeling software systems and it provides standard extension mechanisms based on additional features which can be used to extend its semantic in standard and consistent way. UML profile for schedulability, performance and time specification allows the specification of quantitative information directly in the UML model. [20] Proposed an algorithm for the software performance modeling based on UML as software description notation and multiclass QN [7] as the performance model.

The motivation in this [12] is the use of the Unified Modeling Language (UML) [11], and more specifically the UML Profile for modeling Real Time Systems (RT-UML) profile, as a graphical modeling language for XML Real-Timed Web Services composition and the verification of these systems by using Model Checking techniques on Timed Automata. Some web services workflow patterns and extension in UML to model these patterns are provided in [8]. The patterns identified are Web Service Call, Loop, Data Transformation and Alternate Services. The proposed solutions contain UML activity diagrams along with required extensions. The paper addresses the service composition patterns, but does not provide any feedback mechanism for refinement. [2] talks about MDA approach for development of Web Services and their compositions. Business processes are shown by activity diagrams and static structures by UML class diagrams. Authors have shown mapping from UML to BPEL4WS, WSDL and Java Platform. The UML is a graphical language for visualizing, specifying, constructing, and documenting the artifacts of a software-intensive system [9s]. The UML is rather software-specific and strongly characterized by object orientation. While modeling in systems engineering is interdisciplinary. The use of UML can easily lead to acceptance problems and misunderstandings when in interdisciplinary communication. This led to develop our methodology. All these works are referred to developing UI for software system but not for system engineering. Hence based on [4] we have proposed our methodology.

# 1 Methodology

i. Consider the prototype of a User Interface (UI).

ii. Identify UI elements in the selected prototype

iii. Develop the flow diagram of the UI elements

iv. Develop Activity Model

v. Develop Use Case Model

vi. Refine & Iterate.

The UI elements identified in step 2 are workspaces and functional elements located at the user interface. These are identified by the designer from the Prototype created at Step1. It is mapping from logical and physical interface design to use case descriptions which is the focus of the research reported here.

# 1 Case Study on Amazon.com

## 4.1 Prototype of Amazon.com web service

As an illustration of this methodology, consider the web site Amazon.com, and apply the algorithm for the module Login. The prototype of the login page is given in Fig. 1.

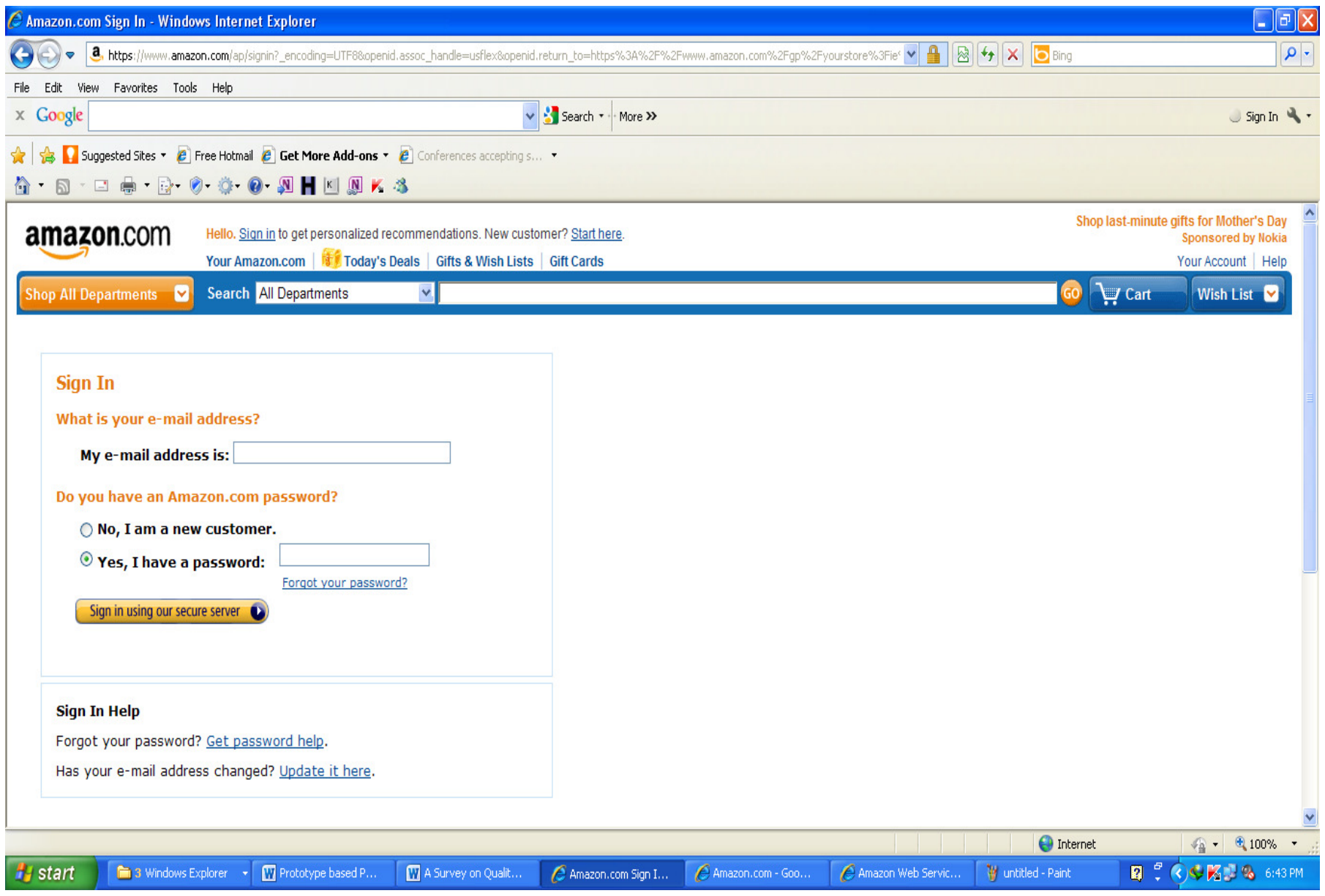

Figure 1. Screen shot of login page of Amazon.com.

## 4.1 UI elements of Login

In Fig. 2, the UI elements column identifies four workspaces (W1- W4) as being required to support the user and system activities. It also identifies two functional elements - Sign in using secure server, Create Account, which are required to support the user tasks. There are two sub-flows in the figure 2, and one exception flow is shown following invalid user input. The information to be displayed is shown in each workspace. This is derived from Prototype model.

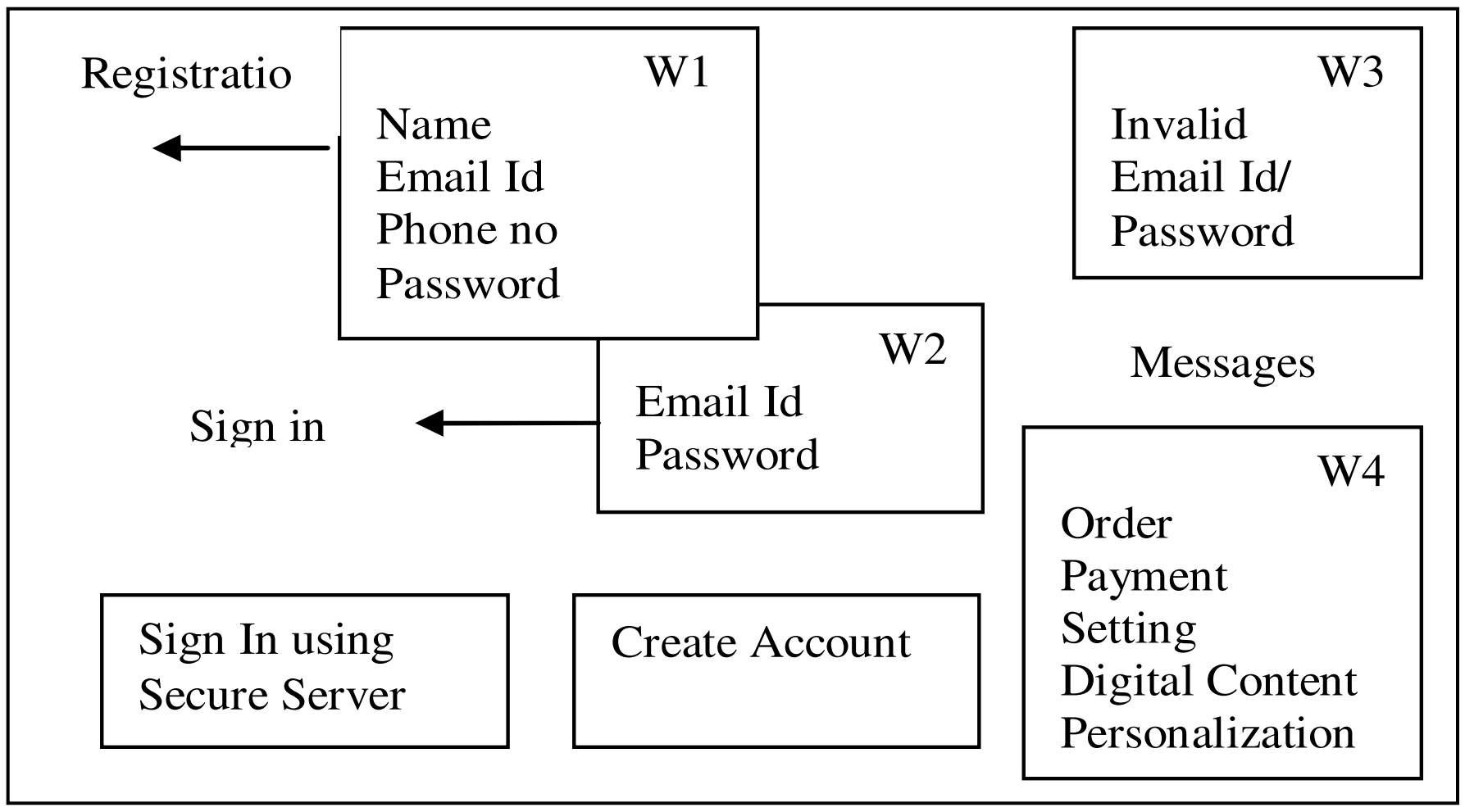

Figure 2. UI Element cluster for the Login use.

## 4.1 Flow diagram of the UI elements

### 4.3.1 Main Flow

In the given below figure 3 showing the 'Main flow of login', where the first workspace W1 showing to create a account by providing the Name, Email id, Phone number and password. When it successfully created then workspace W2 showing how to login, for login we have to provide the Email-id and password. If Email-id and password is correct then it enters in secure server. The last workspace W4 shows entering in to system and can access the functionalities.

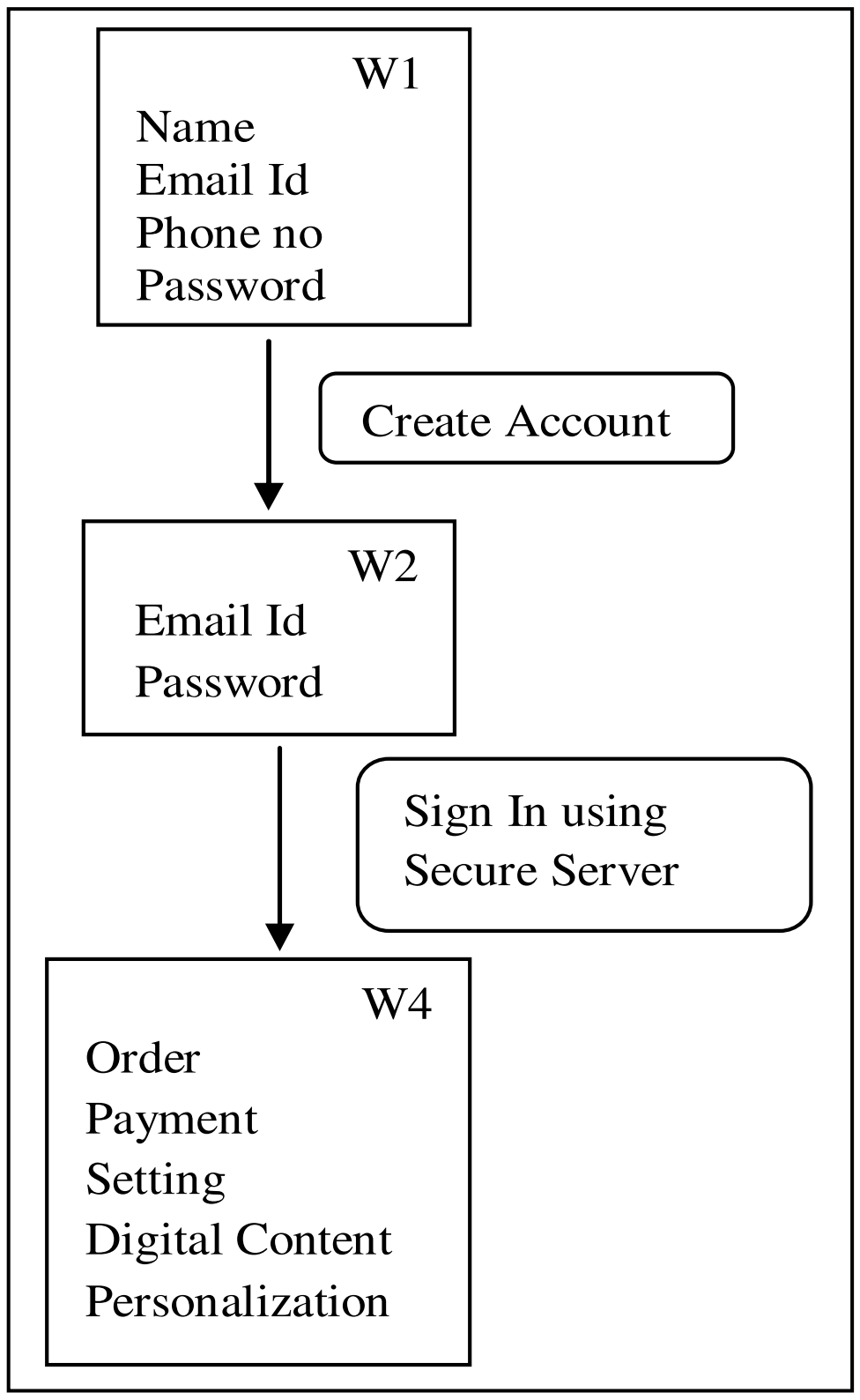

**Figure 3. Main flow of login.**

### 4.3.2 Exception flows

In Fig. 4, the workspace W3 is showing Exception handling when user is logging in. If Email-id or password or both are Incorrect, then it will show the Exception.

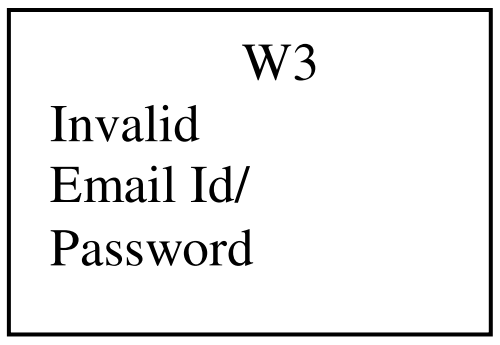

Figure 4. Exception flows of login.

## 4.1 Activity Model for login

In Fig. 5, it is showing the activity model for login. When user wants to enter in to the system, if he is new user then he has to provide the Email-id and fill all the details, then he will enter into the system. It he is the existing user then he has to provide email-id or password. If email-id or password or both is incorrect then display error message and again back to login page. If e-mail-id and password is correct then it moves on to open user form.

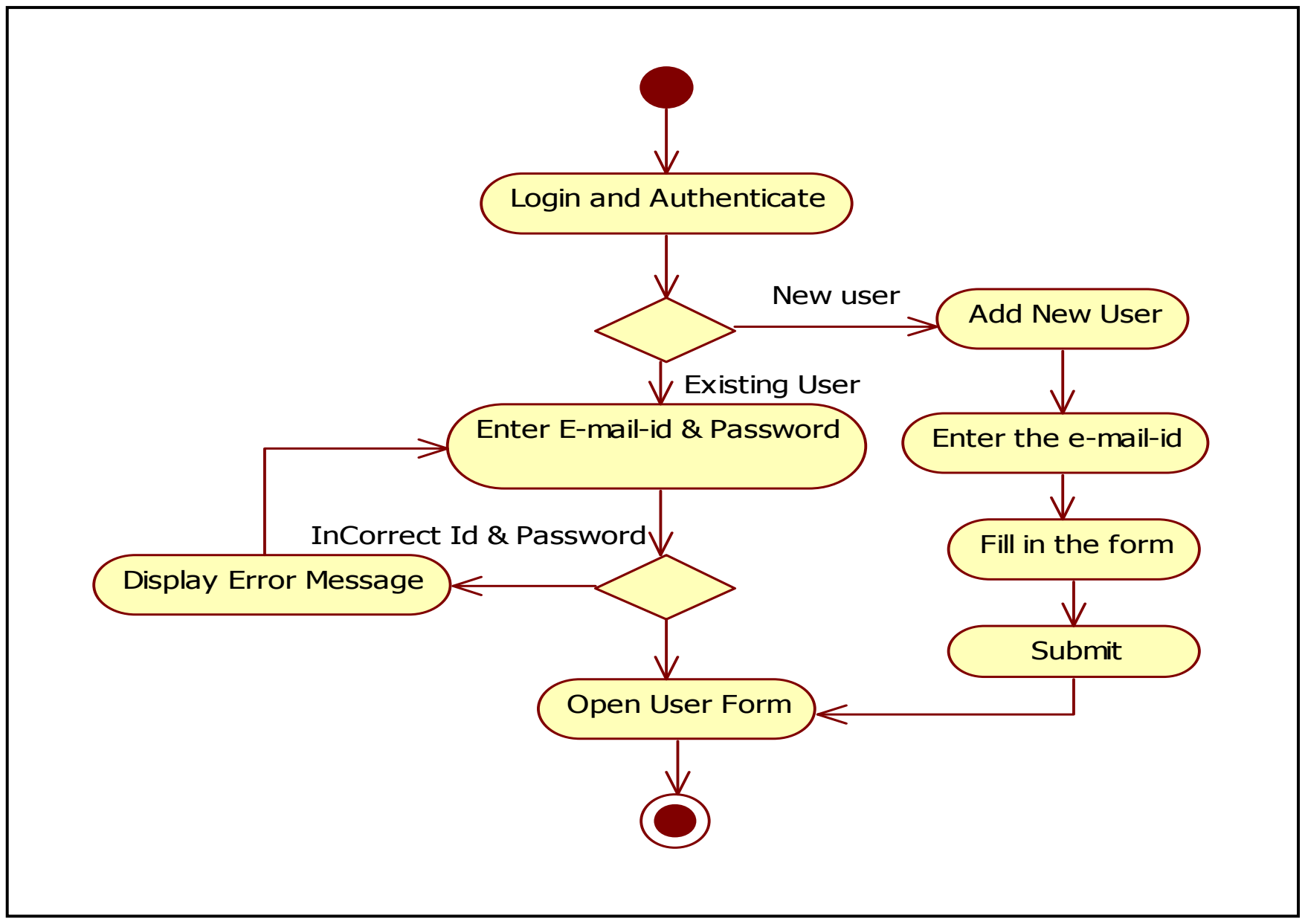

Figure 5. Activity model for login

## 4.1 Use Case Model

In Table 1, it is showing use case specification for login and authenticate. It is showing all the description about the use case as who is actor, pre-conditions, post-conditions, basic steps of login, exception handling.

**Table 1. Use Case Specification – Login and Authenticate.**

| | |
|---|---|
| ***Brief Description*** | This use case describes the process by which a user logs into the System. |
| ***Actor*** | User |
| ***Pre-Conditions*** | **Precondition One** The User must have a valid user name and password. **Precondition Two** The User has access to the system. |
| ***Post-Conditions*** | **Post-Condition One** The User successfully logs into the system and is re-directed to the page. **Post-Condition Two** All and/or any of the login credentials of the User are not valid. The user is informed that the login are invalid and to try again. |
| ***Basic Steps*** | **1.** If new user, the user should enter the credentials and the system redirects the user to the web page.<br>**2.** Already existing user, The user enters and submits his/her Username and password.<br>**3.** The System validates and authenticates the user Information.<br>4. The System re-directs the user to the web page. |
| **Exception Flows** | • The system displays a message to the user indicating that the user name and/or password are incorrect and to try again.<br>• The system displays a message to the user indicating to enter user name and password.<br>• The user continues at basic step #2.<br>• The system displays a message to the user indicating a communication error.<br>• The user exits the system. |

We have obtained the remaining UI Elements and Activity models, the similar procedure as given in figures 6, 7 & Table 2 for Search, figures 8,9,10 & Table 3 for shopping cart, figures 11, 12 & Table 4 for online payment and figures 13, 14 for Wish list respectively.

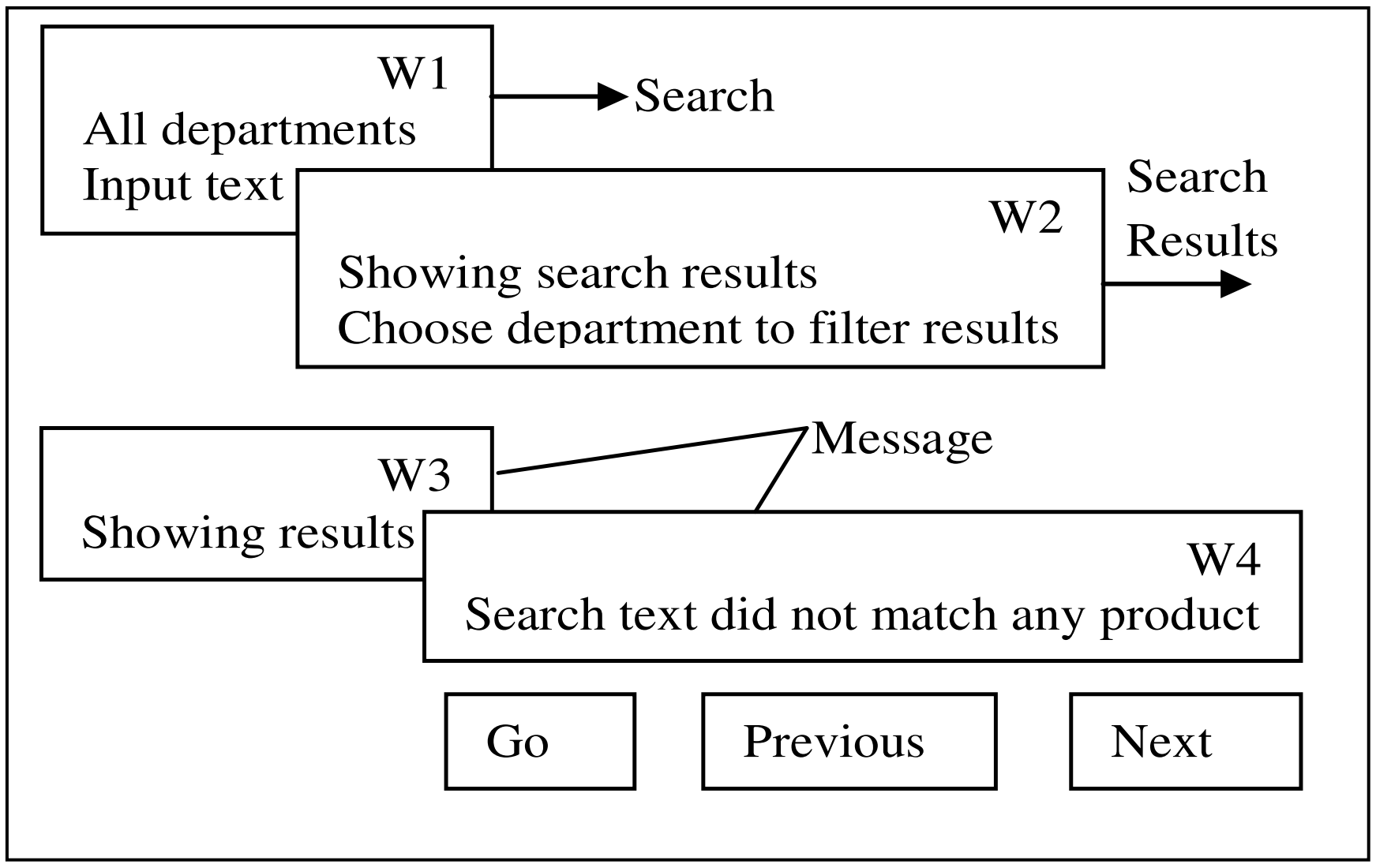

Figure 6. UI Elements Cluster for Search.

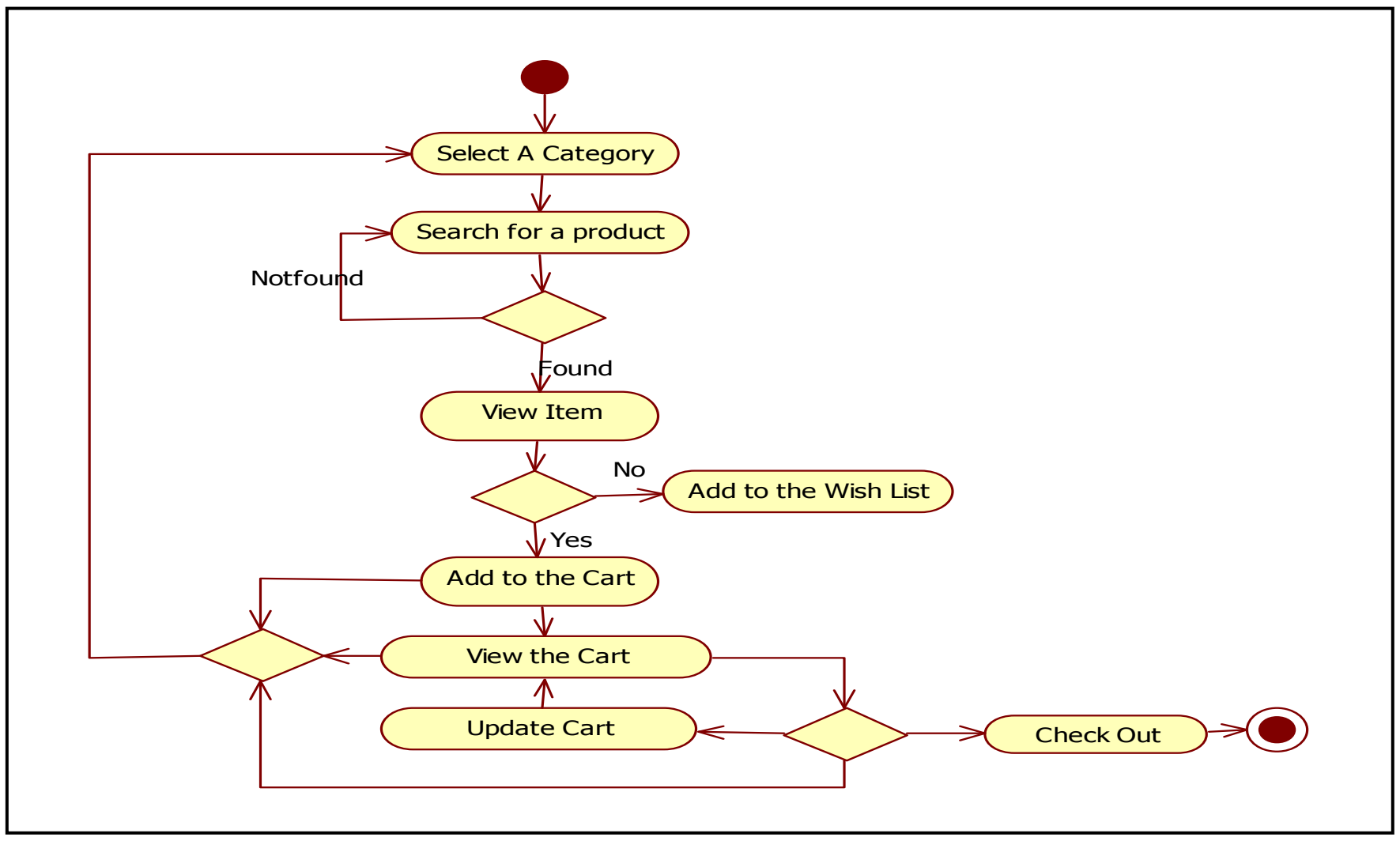

Figure 7. Activity Diagram for Search.

**Table 2. Use Case Specification – Search.**

| | |
|---|---|
| ***Brief Description*** | This use case describes the process by which user can search for product. |
| ***Actor*** | User |
| ***Pre-Condition*** | User must enter the category of search.<br>User must know what product to search for. |
| ***Post -Condition*** | Displays the list of products related to the search.<br>If searched product is not found display the message product not found. |
| ***Basic Steps*** | User must select the product to search.<br>System displays the related product list to the user.<br>View the product and add to the wish list. |
| ***Exception Flows*** | System displays a message to the user indicating “your search does not match any products” |

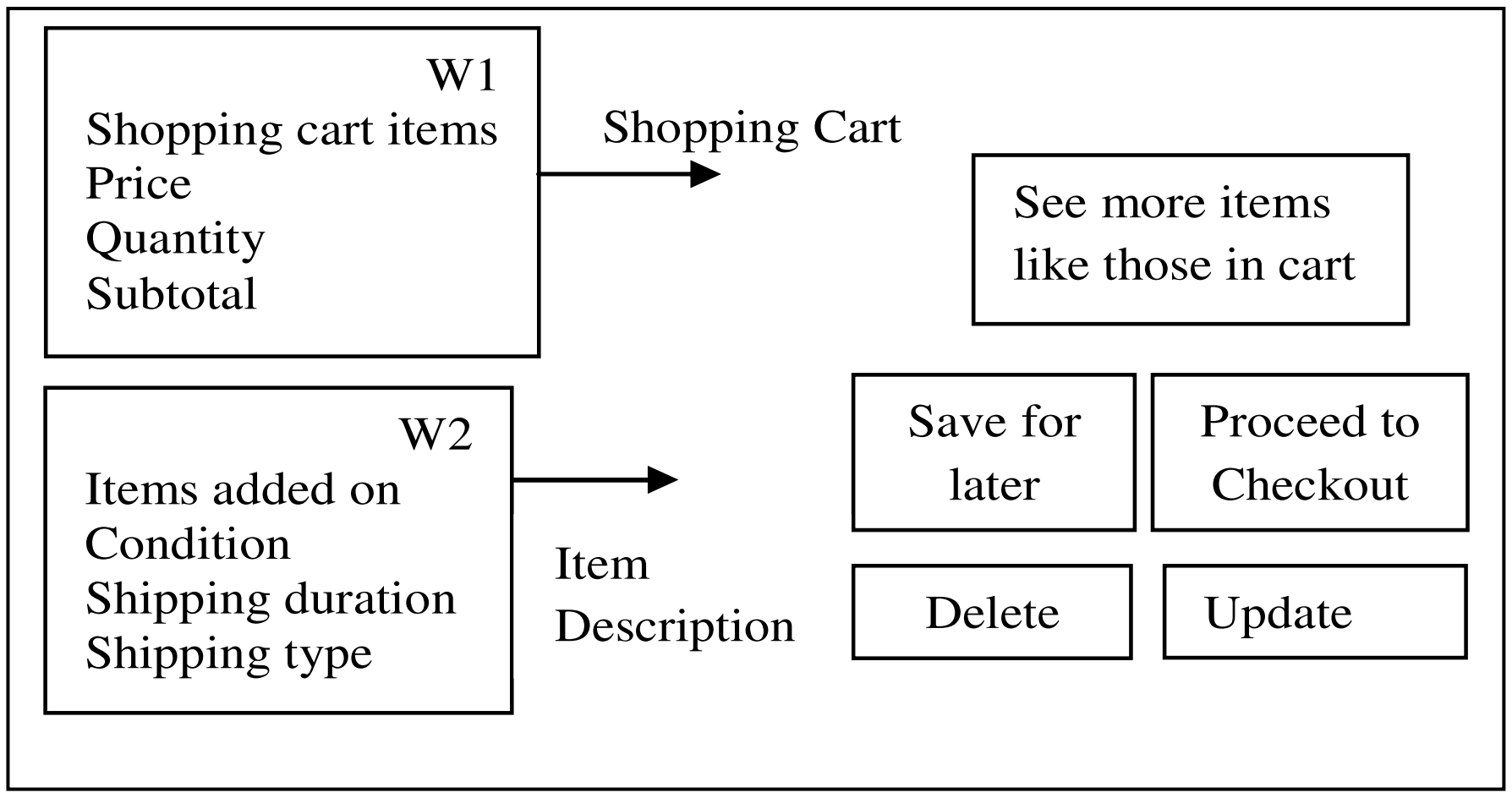

Figure 8. UI Elements Cluster for Shopping Cart.

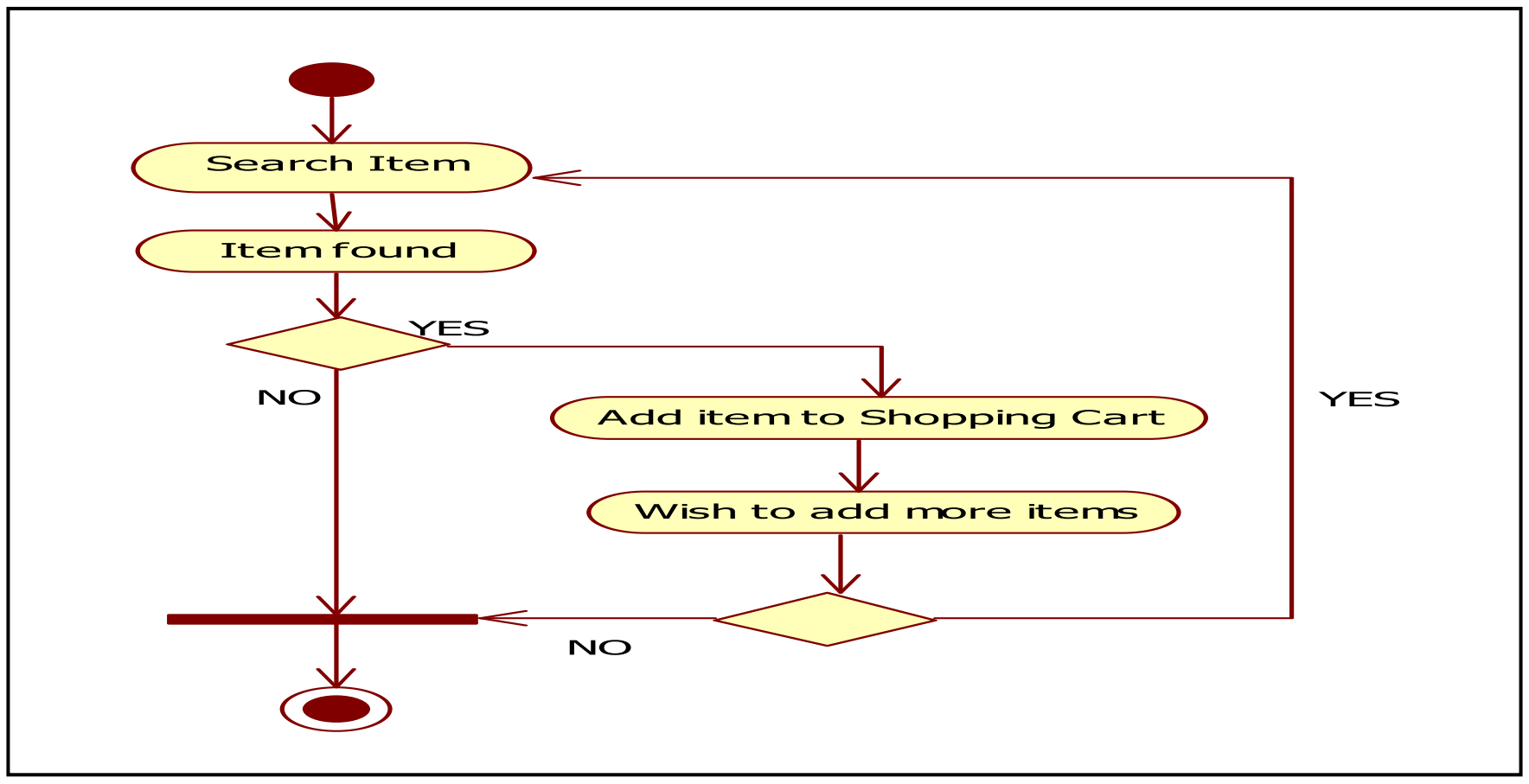

Figure 9. Add Item to the Shopping Cart.

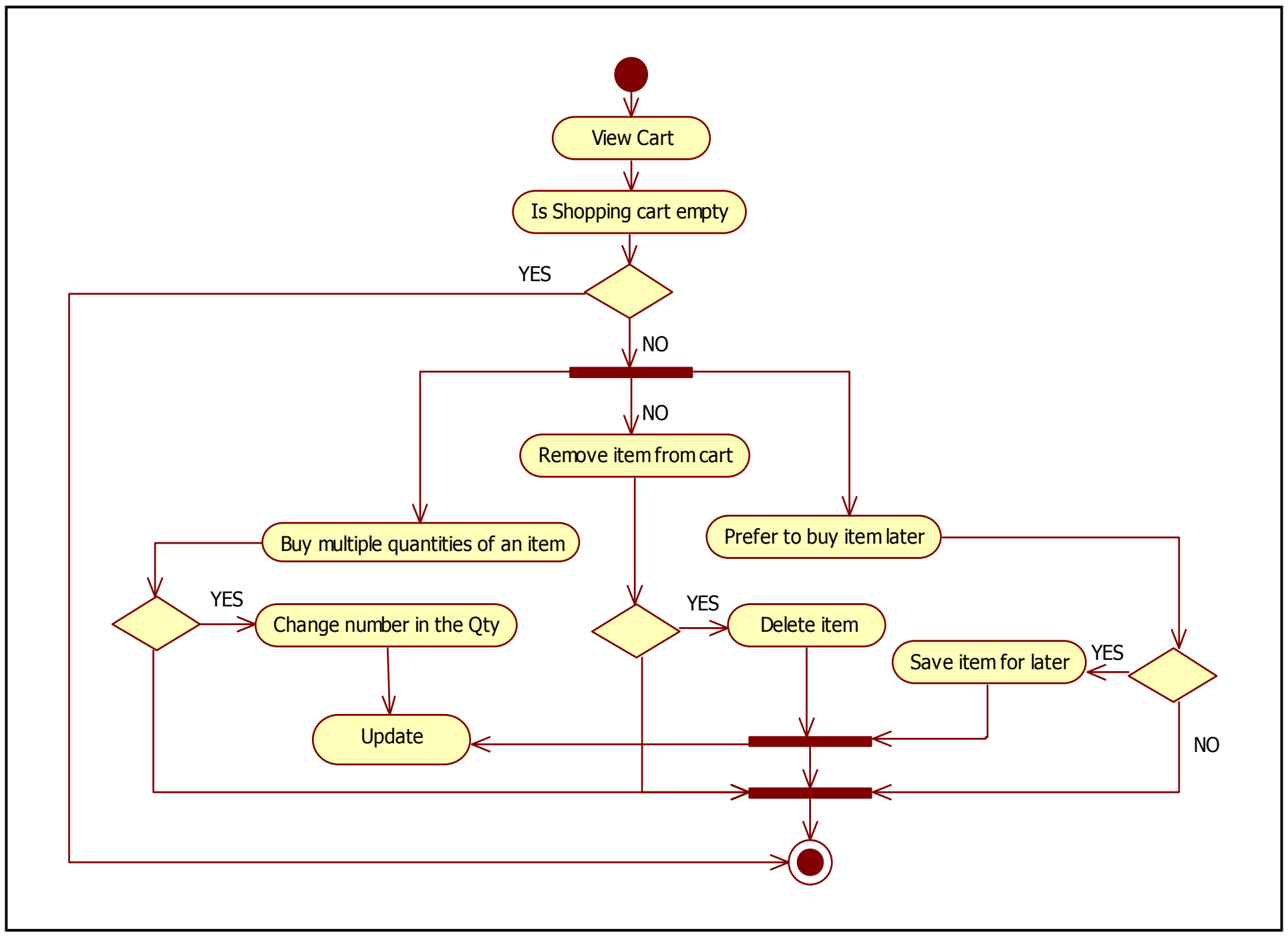

Figure 10. Edit /Remove Item from the Shopping Cart.

**Table 3. Use Case Specification – Shopping Cart.**

| | |
|---|---|
| ***Brief Description*** | This use case describes the process by which user can edit/remove the item form the cart. |
| ***Actor*** | User |
| ***Pre-Condition*** | User must be authorized to edit cart.<br>User must be logged into the system.<br>User must check for the cart is empty or not.<br>User must know what to edit in the cart. |
| ***Post -Condition*** | Cart will be edited with the changes required by the user. |
| ***Basic Steps*** | User must be authenticated to edit the cart.<br>User must check for shipping cart is empty or not.<br>If cart is not empty user must select the preferred action to be taken (Edit/Remove) on the cart.<br>Action will be performed by the system and communicates the message to the user. |
| ***Exception Flows*** | If the cart is empty displays the message that the cart is empty.<br>Redirect to the login webpage if user had not logged into the account. |

Figure 11. UI Elements Cluster for Online Payment.

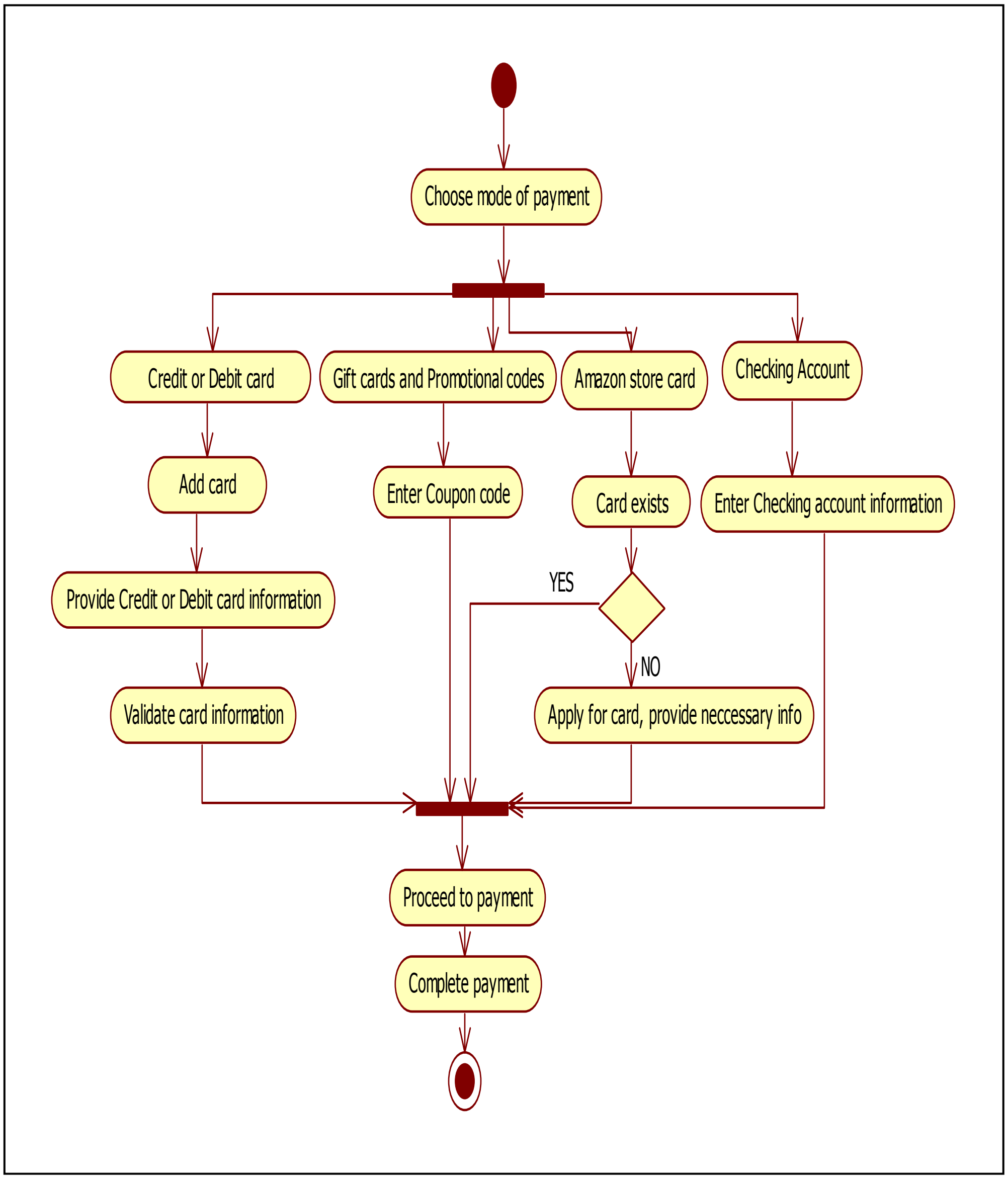

Figure 12. Activity Diagram for Online Payment.

**Table 4. Use Case Specification – Online Payment.**

| | |
|---|---|
| ***Brief Description*** | This use case describes the process by which user can do the online payment. |
| ***Actor*** | User |
| ***Pre-Condition*** | User must be authorized to do the online payment.<br>User must register the card for the payment.<br>User must choose the mode of payment.<br>User must choose the type of the card |
| ***Post -Condition*** | System display the message payment is successful if payment is success. |
| | System displays the message unable to proceed if payment is unsuccessful. |
| ***Basic Steps*** | User must be added the product to the shipping cart.<br>User must choose mode of payment.<br>User should enter the valid card information.<br>Enter the amount to be paid.<br>Redirect to the banking website if information provided by the user is correct.<br>Payment will complete. |
| ***Exception Flows*** | Displays the communication information if the details entered by the user are incorrect.<br>Redirect to the login webpage if user had not logged into the account. |

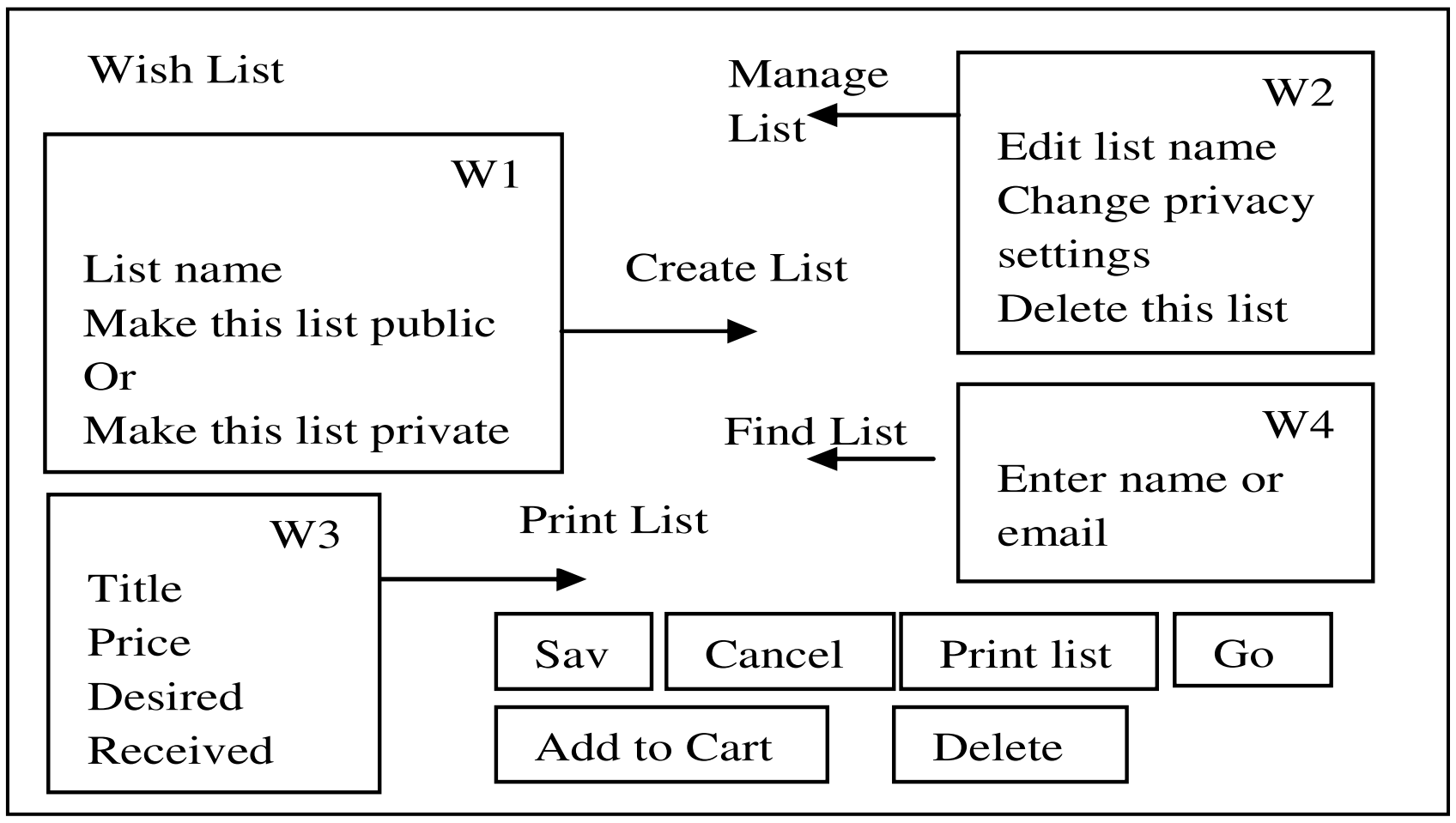

Figure 13. UI Elements Cluster of Wish List.

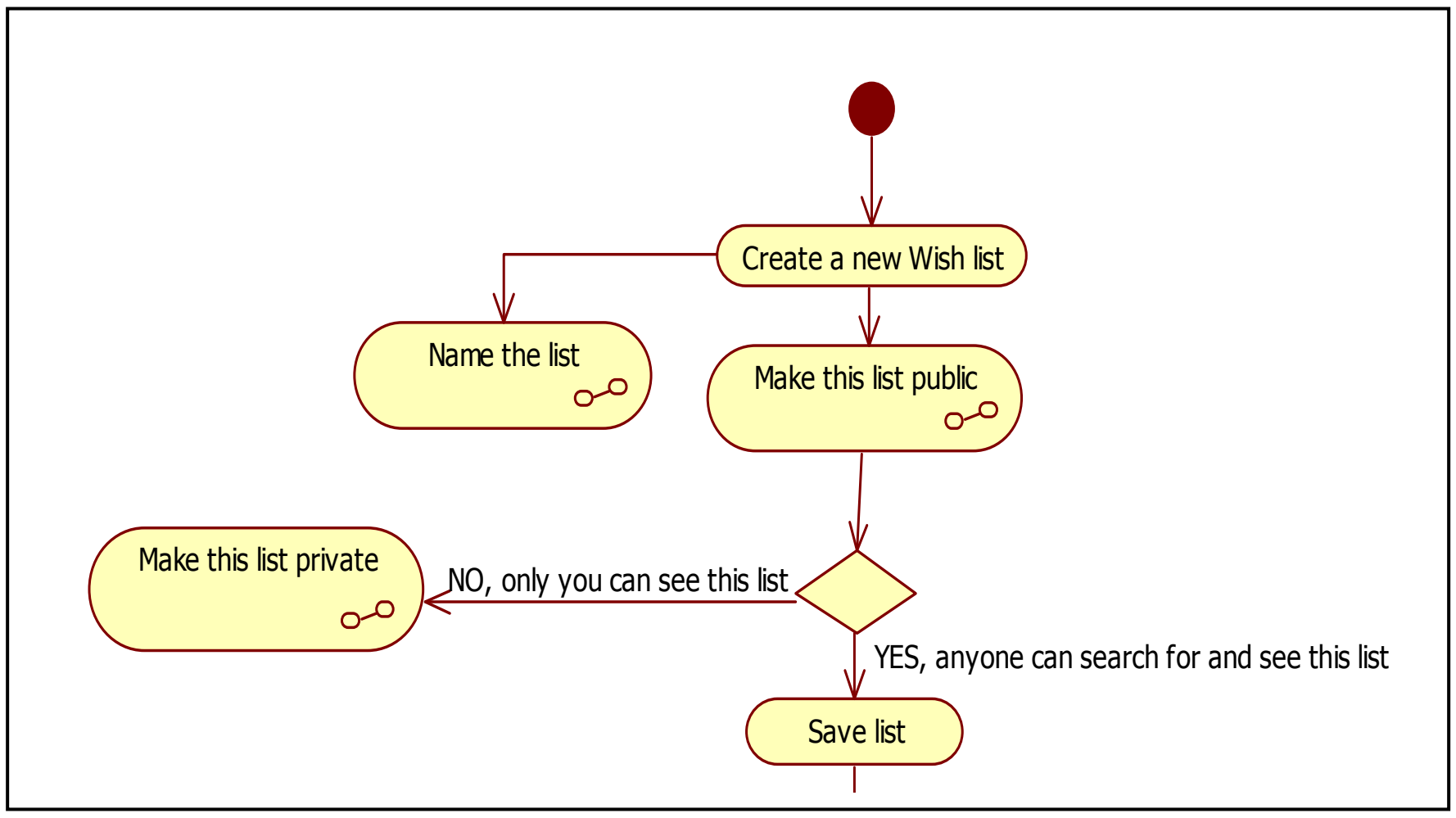

Figure 14. Activity Diagram of Wish List.

**Table 5. Use Case Specification – Wish List.**

| | |
|---|---|
| ***Brief Description*** | This use case describes the process by which user can add products to the wish list. |
| ***Actor*** | User |
| ***Pre-Condition*** | User must be authorized to add the items to the wish list.<br>User must select the items to be added to the wish list.<br>User need to decide whether to make the list as public or private.<br>User need to create the list if list does not exist already. |
| ***Post -Condition*** | List of items will be added to the wish list. |
| ***Basic Steps*** | User must be authenticated to enter the items to the wish list.<br>Select the item need to be added to the wish list.<br>Make the list as private/public.<br>Add the items to the list.<br>Save the list. |
| ***Exception Flows*** | Redirect to the login webpage if user had not logged into the account |

# 4 Conclusion and Future work

In general, developing prototype for software applications is industry practice. Various methodologies facilitate to predict performance from UML models. Hence, in this paper, we have proposed a methodology to transform a prototype of UI into UML models, Activity diagram and Use Case diagram. The user interface designer can predict the performance of a software system at the early stages. As future work, we propose to develop methodologies to analyze performance of Web Services from these UML models.